\documentclass[aps,prb,twocolumn]{revtex4} 

\usepackage{graphicx}
\usepackage{dcolumn}
\usepackage{bm}
\usepackage{amsmath}  

\newcommand{\comment}[1]{}

\begin{document}
\renewcommand{\theequation}{\arabic{section}.\arabic{equation}}

\title{Hamilton's Equations of Motion from Schr\"odinger's Equation}


\author{Phil Attard}
\affiliation{ {\tt phil.attard1@gmail.com} } 


\begin{abstract}
Starting from Schr\"odinger's equation,
Hamilton's classical equations of motion
emerge from the collapse of the unsymmetrized wave function
in a decoherent open quantum system entangled with its environment.
\end{abstract}

\pacs{}

\maketitle

%
%

\renewcommand{\theequation}{\arabic{equation}}

\subsubsection{The Ehrenfest Theorem}

Hamilton's equations,
which are equivalent to Newton's equations of motion,
govern the time evolution of classical systems.
It is therefore of some interest to show
how they result from the principles of quantum mechanics,
since this would shed light
on the origins and nature of the classical universe.

The textbook derivation of Hamilton's equations
invokes Ehrenfest's theorem.
However, as the following critical analysis shows,
one can legitimately question whether the theorem
is really relevant to Hamilton's equations,
and whether it actually provides any insight into their basis
or into the nature of the transition from quantum to classical motion.

Recall that Hamilton's equations of motion are
\begin{equation}
\dot{\bf q} =
\frac{\partial {\cal H}({\bf q},{\bf p}) }{\partial {\bf p}},
\mbox{ and }
\dot{\bf p} =
\frac{-\partial {\cal H}({\bf q},{\bf p}) }{\partial {\bf q}} .
\end{equation}
Here the positions of the $N$ particles are
$ {\bf q}
= \{{\bf q}_1,{\bf q}_2, \ldots, {\bf q}_N \}$,
where the position of particle $j$ is
$ {\bf q}_j = \{ q_{jx},q_{jy},q_{jz} \} $,
and similarly for the momenta ${\bf p}$.
The classical Hamiltonian is
the sum of the kinetic and potential energies,
${\cal H}({\bf q},{\bf p}) = {\cal K}({\bf p}) + U({\bf q})$,
where the kinetic energy is
${\cal K}({\bf p})  = {\bf p}\cdot{\bf p}/2m$.

In the present context there are two important points to note:
First, Hamilton's equations use the positions and momenta
of the particles in the system simultaneously
to give their velocities  and accelerations at a given time.
And second, the particles consequently follow a trajectory over time.
Neither of these is compatible with the derivation
via the Ehrenfest theorem, as is now discussed

The conventional interpretation of quantum mechanics
gives the positions and momenta of the particles of the system
in the normalized wave state $\psi$ as expectation values,
\begin{equation}
{\bf q}(\psi)
=
\langle \psi |\hat{\bf r}|\psi\rangle ,
\mbox{ and }
{\bf p}(\psi)
=
\langle \psi |\hat{\bf p}|\psi\rangle .
\end{equation}
The position operator is $\hat{\bf r}={\bf r}$ (or $\hat{\bf q}={\bf q}$),
and the momentum operator is
$\hat{\bf p}({\bf r}) = -\mathrm{i} \hbar \nabla_{\bf r}$.
Now use the facts that the Hamiltonian operator,
$\hat{\cal H}({\bf r}) = {\cal H}({\bf r},\hat{\bf p})$,
is Hermitian,
and that the Schr\"odinger equation is
$\mathrm{i}\hbar\dot \psi =  \hat{\cal H}  \psi $,
where $\hbar$ is Planck's constant divided by $2\pi$.
With these the rate of change of position is
\begin{eqnarray}
\frac{\partial {\bf q}(\psi) }{\partial t}
& = &
\langle \dot\psi |\hat{\bf q}|\psi\rangle
+
\langle \psi |\hat{\bf q}| \dot\psi \rangle
\nonumber \\ & = &
\langle \frac{1}{\mathrm{i}\hbar} \hat{\cal H}\psi |\hat{\bf q}|\psi\rangle
+
\langle \psi |\hat{\bf q}| \frac{1}{\mathrm{i}\hbar} \hat{\cal H} \psi\rangle
\nonumber \\ & = &
\frac{-1}{\mathrm{i}\hbar}
\langle \psi |  \hat{\cal H} \hat{\bf q}|\psi\rangle
+
\frac{1}{\mathrm{i}\hbar}
\langle \psi |  \hat{\bf q} \hat{\cal H} |\psi\rangle
\nonumber \\ & = &
\frac{-1}{2m\mathrm{i}\hbar}
\langle \psi |  \hat{\bf p}^2 {\bf r}|\psi\rangle
+
\frac{1}{2m\mathrm{i}\hbar}
\langle \psi |  {\bf r} \hat{\bf p}^2 |\psi\rangle
\nonumber \\ & = &
\frac{-2}{2m\mathrm{i}\hbar}
\langle \psi |  (\hat{\bf p}  {\bf r}) \cdot \hat{\bf p} |\psi\rangle
\nonumber \\ & = &
\frac{1}{m}
\langle \psi |  \hat{\bf p} |\psi\rangle
\nonumber \\ & = &
\langle \psi  |
\frac{\partial \hat{\cal H} }{\partial \hat{\bf p}}
|\psi \rangle .
\end{eqnarray}
Similarly the rate of change of momentum is
\begin{eqnarray}
\frac{\partial {\bf p}(\psi) }{\partial t}
& = &
\langle \dot\psi |\hat{\bf p}|\psi\rangle
+
\langle \psi |\hat{\bf p}| \dot\psi \rangle
\nonumber \\ & = &
\langle \frac{1}{\mathrm{i}\hbar} \hat{\cal H}\psi |\hat{\bf p}|\psi\rangle
+
\langle \psi |\hat{\bf p}| \frac{1}{\mathrm{i}\hbar} \hat{\cal H} \psi\rangle
\nonumber \\ & = &
\frac{-1}{\mathrm{i}\hbar}
\langle \psi |  \hat{\cal H} \hat{\bf p}|\psi\rangle
+
\frac{1}{\mathrm{i}\hbar}
\langle \psi |  \hat{\bf p} \hat{\cal H} |\psi\rangle
\nonumber \\ & = &
\frac{-1}{\mathrm{i}\hbar}
\langle \psi | U({\bf r}) \hat{\bf p}|\psi\rangle
+
\frac{1}{\mathrm{i}\hbar}
\langle \psi |  \hat{\bf p} U({\bf r})  |\psi\rangle
\nonumber \\ & = &
\langle \psi |  (-\nabla U({\bf r}))  |\psi\rangle
\nonumber \\ & = &
\langle \psi  |
\frac{-\partial \hat{\cal H} }{\partial \hat{\bf q}}
|\psi \rangle .
\end{eqnarray}
These constitute the Ehrenfest theorem,
and many believe that the resemblance to Hamilton's equations of motion
constitutes a derivation of them.

The contentious issue for the Ehrenfest theorem
lies with the physical interpretation
rather than with the mathematical manipulations.
It is questionable whether the implied physical content is consistent
with the standard interpretation of quantum mechanics,
or whether these expressions in any sense explain
the physical origins of classical mechanics.

These problems
are connected with the non-commutativity of the position and momentum
operators.
Since these cannot be applied simultaneously,
it is not possible to interpret the expectation values
${\bf q}(\psi)$ and ${\bf p}(\psi)$
as the position and momentum of the particles in the system
at a given time.
Further, since an expectation value
is commonly interpreted as a measurement,
and since a measurement is commonly taken to perturb the
wave function of the system,
the above expectation values cannot apply to a single wave state $\psi$.

It is a tenet of the Copenhagen interpretation of quantum mechanics
that the position or else the momentum  of a particle
are only realized at the time of measurement or observation,
and that these have no objective reality between such times.
Adherents to these views believe
that particles do not follow real trajectories,
which is difficult to reconcile
with the existence of particle trajectories in classical mechanics.
Faithful followers of Bohr
invoke the quantum mechanical Ehrenfest theorem
as the explanation of classical behavior,
while disavowing the physical trajectories that result.

A further issue that casts doubt on the Ehrenfest theorem as an
explanation of classical mechanics
is that the wave function $\psi$ that is used
can validly be expressed as a linear superposition of
the allowed states of the system.
But Hamilton's equations are non-linear functions
of the position and momentum states of the system.
There is nothing in the Ehrenfest theorem  that explains
why the superposition of states does not occur in the classical world.

Of course there is no question that Hamilton's equations of motion
apply in the classical regime.
The issues are:
Can one  convincingly derive those equations
from the principles of quantum mechanics?
What is the physical interpretation of such a derivation?
And how does this explain the origins of the classical world?

\subsubsection{Hamilton's Equations of Motion}

A more realistic derivation of Hamilton's equations is as follows.
Consider an isolated system
that consists of a subsystem of primary interest
containing  $N$ particles with positions ${\bf q}$ (or ${\bf r}$)
and momenta ${\bf p}$,
and a reservoir (environment) with which
conserved quantities such as energy or momentum can be exchanged.
Two assumptions are required:
The interactions with the reservoir are strong enough
to entangle the total wave function,
which causes the subsystem wave function to collapse
into the principle states of the exchanged variable.
And those same interactions are weak enough
that they perturb negligibly
the adiabatic time evolution of the subsystem.

Entanglement means that the subsystem consists
of a mixture of pure  states of the exchangeable variable
rather than a superposition of states,
so that its wave function is decoherent in those states.
Environment-induced decoherency
has been analysed in the context of quantum measurement theory
(Zeh 2001, Zurek 2003, Schlosshauer  2005)
and in the context of quantum statistical mechanics
(Attard 2015, 2018, 2021, 2023a).

The position and momentum eigenfunctions of the subsystem are
\begin{eqnarray}
| {\bf q}\rangle
&\equiv &
\zeta_{\bf q}({\bf r})
 = \delta( {\bf q}-{\bf r}),
\nonumber \\ \mbox{ and  }
| {\bf p}\rangle
& \equiv &
\zeta_{\bf p}({\bf r})
= V^{-N/2} e^{-{\bf p}\cdot {\bf r}/\mathrm{i}\hbar} .
\end{eqnarray}
Because the position belongs to the continuum,
the normalization is a Dirac-$\delta$ function,
$ \langle {\bf q}'| {\bf q}\rangle = \delta( {\bf q}-{\bf q}')$
(Messiah 1961, Merzbacher  1970).
The momenta are quantized, $ {\bf p} = {\bf n} \Delta_p $
(${\bf n}$ is a $3N$-dimensional vector of integers)
with spacing $\Delta_p =2\pi\hbar/L$,
where the volume of the cubic subsystem is $V=L^3$
(Messiah 1961, Merzbacher  1970).
A phase space point is denoted
${\bf \Gamma} = \{ {\bf q} ,{\bf p} \}$.

\begin{subequations}
The evolution of a momentum eigenfunction
over a time interval $\tau$
may be written in two possible ways,
\begin{eqnarray} \label{Eq:axiom+}
\hat U^0({\bf q};\tau) \zeta_{\bf p}({\bf q}) 
& = &
\sum_{{\bf p}'}
\langle  {\bf p}'| \hat U^0({\bf r}';\tau) | {\bf p}\rangle
\, \zeta_{{\bf p}'}({\bf q})
\nonumber \\ & = &
 \zeta_{\overline{\bf p}'}(\overline{\bf q}').
\end{eqnarray}
The first equality
is a formal expansion in terms of the momentum basis function,
and would be standard for an isolated subsystem.
The second equality is a significant statement
about the decoherent nature of the subsystem.
This  says  that given an initial momentum state and position,
there is a single most likely destination momentum state and position.
As is discussed below,
this only applies in a regime where wave function symmetrization effects
are negligible.
An expression for
$\overline{\bf \Gamma}'=\overline{\bf \Gamma}(\tau|{\bf \Gamma})$
is  obtained next.

The axiom for the backward transition is
\begin{eqnarray} \label{Eq:axiom-}
\hat U^0(\overline{\bf q}';\tau)^\dag
\zeta_{\overline{\bf p}'}(\overline{\bf q}')
& = &
\sum_{{\bf p}''}
\langle  {\bf p}''| \hat U^0({\bf r}';\tau)^\dag | \overline{\bf p}'\rangle
\, \zeta_{{\bf p}''}(\overline{\bf q}')
\nonumber \\ & = &
 \zeta_{{\bf p}}({\bf q}).
\end{eqnarray}
This is a statement that the motion is reversible.
The argument of the conjugate propagator 
is $\overline{\bf q}'$, not ${\bf q}$,
which means that it contains more information
than simply inverting  the first  equation.
The two differ by a term ${\cal O}(\tau)$,
which vanishes upon application of the following result.
\end{subequations}

For a subsystem dominated by adiabatic evolution,
the time propagator is
$ \hat U^0(\tau) = e^{\tau \hat {\cal H}/\mathrm{i}\hbar}
= 1 + \tau \hat {\cal H}/\mathrm{i}\hbar + {\cal O}(\tau^2)$.
The momentum eigenfunction is an eigenfunction
of the kinetic energy operator,
which means that  $\hat{\cal H}({\bf q}) \zeta_{{\bf p}}({\bf q})
= {\cal H}({\bf q},\hat{\bf p}) \zeta_{{\bf p}}({\bf q})
= \zeta_{{\bf p}}({\bf q}) {\cal H}({\bf q},{\bf p})$.
Hence the product of the left hand sides of the two axioms  is
\begin{eqnarray}
\lefteqn{
\big\{ \hat U^0({\bf q};\tau)   \zeta_{{\bf p}}({\bf q}) \big\}
\big\{ \hat U^0(\overline{\bf q}';\tau)^\dag
\zeta_{\overline{\bf p}'}(\overline{\bf q}') \big\}
} \nonumber \\
& = &
\zeta_{{\bf p}}({\bf q})
\Big\{ 1 + \frac{\tau}{\mathrm{i}\hbar} {\cal H}({\bf q},{\bf p})
+{\cal O}(\tau^2) \Big\}
\nonumber \\ && \mbox{ } \times
\zeta_{\overline{\bf p}'}(\overline{\bf q}')
\Big\{ 1 - \frac{\tau}{\mathrm{i}\hbar}
{\cal H}(\overline{\bf q}',\overline{\bf p}')
+{\cal O}(\tau^2) \Big\}
\nonumber\\ & = &
 \zeta_{{\bf p}}({\bf q})\zeta_{\overline{\bf p}'}(\overline{\bf q}')
\Big\{ 1
- \frac{\tau}{\mathrm{i}\hbar}
\Delta_{\bf q}^0 \cdot \nabla_q {\cal H}({\bf q},{\bf p})
\nonumber \\ && \mbox{ }
- \frac{\tau}{\mathrm{i}\hbar}
\Delta_{\bf p}^0 \cdot \nabla_p {\cal H}({\bf q},{\bf p})
+{\cal O}(\tau^3) \Big\} .
\end{eqnarray}
Here $\Delta_{\bf q}^0 = \overline{\bf q}'-{\bf q} = {\cal O}(\tau) $
and  $\Delta_{\bf p}^0 = \overline{\bf p}'-{\bf p} = {\cal O}(\tau) $.
Note that the remaining terms that are ${\cal O}(\tau^2)$ cancel.

Since the  pre-factor equals the product of the right hand sides,
the embraced terms must sum to unity.
Hence the change in position must be proportional to
the momentum gradient of the  Hamiltonian,
and the change in momentum  must be proportional to
the negative of the position gradient of the  Hamiltonian.
Dimensional considerations
show that the common  proportionality factor is just the time step,
and so one must have that
\begin{eqnarray}
\dot{\bf q}^0
& \equiv &
\lim_{\tau\to 0} \frac{\Delta_{\bf q}^0}{\tau}
= \nabla_p {\cal H}({\bf q},{\bf p}),
\nonumber \\
\mbox{ and }
\dot{\bf p}^0
& \equiv & \lim_{\tau\to 0} \frac{\Delta_{\bf p}^0}{\tau}
= - \nabla_q {\cal H}({\bf q},{\bf p}) .
\end{eqnarray}
These are just Hamilton's classical equations of motion.

\subsubsection{Discussion}

The main result in this paper
is that classical behavior occurs,
and only occurs, in an open quantum system
with negligible wave function symmetrization.
The derivation assumed that the interactions
with the reservoir or environment
entangled and collapsed the subsystem wave function
without significantly  perturbing Schr\"odinger's equation
for its rate of change.

The direct effects of the reservoir or environment
on the time evolution of the subsystem
are briefly discussed in appendix~\ref{Sec:SDE},
where the explicit form follows from
thermodynamic principles and probability theory.
This procedure has been used for the development
of stochastic molecular dynamics for classical systems
(Attard 2012, 2021, 2023a).
It has also been recently used to derive equations of motion
and to calculate the viscosity of superfluid helium (Attard 2023b).

The justification for the axioms (\ref{Eq:axiom+}) and (\ref{Eq:axiom-})
is that in an open quantum system the wave function is decoherent
in the exchangeable states,
which means that it has collapsed into
a mixture of pure states rather than a superposition of states
(Zeh 2001, Zurek 2003, Schlosshauer  2005, Attard 2015, 2018, 2021, 2023a).
The right hand side of the first equality in these equations
is for a position in a superposition of momentum states,
whereas that of the second equality
is for a position in a pure momentum state.

The result only applies in regimes
in which wave function symmetrization is negligible,
as follows from the  non-linear nature of the expressions.
The evolution of the superposition of two distinct momentum states
has no meaning in the scheme,
\begin{eqnarray}
\lefteqn{
\frac{1}{\surd 2} \hat U^0({\bf q};\tau)
\big[ \zeta_{{\bf p}^{(1)}}({\bf q}) + \zeta_{{\bf p}^{(2)}}({\bf q}) \big]
} \nonumber \\
& \ne &
\frac{1}{\surd 2}
\big[ \zeta_{\overline{\bf p}'^{(1)}}(\overline{\bf q}')
+
\zeta_{\overline{\bf p}'^{(2)}}(\overline{\bf q}') \big],
\;\;
{\bf p}^{(1)} \ne {\bf p}^{(2)}.
\end{eqnarray}
Since the symmetrization or anti-symmetrization of the wave function
corresponds to such a superposition of permuted states,
it is clear that  the present mechanism does not hold
for a  symmetrized wave function.

For an isolated quantum system,
the permutation operator and the Hamiltonian time propagator commute,
which conserves   wave function symmetrization over time
(Messiah 1961, Merzbacher  1970).
In the present case,
the permutation operator and the classical Hamiltonian evolution
do not commute.
In the context of the classical phase space formulation
of quantum statistical mechanics (Attard 2015, 2018, 2021, 2023a),
whereas the symmetrization factor is constant
under Schr\"odinger's equation,
the symmetrization function of classical phase space
is not constant under Hamilton's equations of motion
(except when it is unity).
More generally, one concludes that classical Hamiltonian evolution
is only valid when symmetrization can be neglected
(i.e.\ the symmetrization function is unity),
as is the case when the momentum states are either empty or singly occupied,
which occurs in the high temperature or low density regime.

Conversely, when wave function symmetrization is not negligible,
as in Bose-Einstein condensation,
the evolution of an open quantum system is not dictated
by Hamilton's equations of motion.
This justifies the development of non-Hamiltonian equations
for the motion of condensed bosons,
which have been used to explain
the physical basis of superfluid flow (Attard 2023b).

Commonly,  superfluidity is thought to be surprising
because  it represents fluid flow without viscosity.
However,
since the basis to understanding viscosity is the transfer of momentum
in classical collisions, it is not realistic to expect normal viscosity
in a condensed boson system in which classical collisions do not occur.
The recently discovered side-step collisions
that circumvent momentum transfer between condensed bosons
certainly violate classical intuition (Attard 2023b).
However, since one conclusion from
the present results is that Hamiltonian dynamics
cannot determine the motion of condensed bosons,
one must accept that
classical intuition is not a reliable guide to the nature of superfluid flow.


The present derivation of Hamilton's
equations of motion 
shows that they arise from the collapse of the wave function
due to entanglement with the reservoir or environment,
and that they apply in the regime of negligible wave function symmetrization.
These are ultimately the origin of classical behavior,
which points are missing from the Ehrenfest theorem.

%
\renewcommand{\theequation}{\arabic{equation}}
%

\section*{References}


\begin{list}{}{\itemindent=-0.5cm \parsep=.5mm \itemsep=.5mm}

\item
Attard  P 2012
\emph{Non-Equilibrium Thermodynamics and Statistical Mechanics:
Foundations and Applications'}
(Oxford: Oxford University Press)

\item
Attard  P 2015
\emph{Quantum Statistical Mechanics:
Equilibrium and Non-Equilibrium Theory from First Principles}
(Bristol: IOP Publishing)

\item
Attard  P 2018
Quantum Statistical Mechanics in Classical Phase Space. Expressions for
  the Multi-Particle Density, the Average Energy, and the Virial Pressure
arXiv:1811.00730

\item 
Attard P  2021
\emph{Quantum Statistical Mechanics in Classical Phase Space}
(Bristol: IOP Publishing)

\item 
Attard  P 2023a
\emph{Entropy Beyond the Second Law.
Thermodynamics and Statistical Mechanics
for Equilibrium, Non-Equilibrium, Classical, And Quantum Systems}
(Bristol: IOP Publishing, 2nd edition)

\item 
Attard  P 2023b
Quantum Stochastic Molecular Dynamics Simulations
of the Viscosity of Superfluid Helium
arXiv:2306.07538

\item 
Merzbacher E 1970
\emph{Quantum Mechanics} 2nd edn
(New York: Wiley)

\item 
Messiah A 1961 \emph{Quantum Mechanics}
(Vol 1 and 2) (Amsterdam: North-Holland)

\item  
Schlosshauer M 2005
Decoherence, the measurement problem, and interpretations of quantum
mechanics
arXiv:quant-ph/0312059v4

\item  
Zeh HD 2001
\emph{The Physical Basis of the Direction of Time} 4th edn
(Berlin: Springer)

\item  
Zurek WH 2003
Decoherence, einselection,
and the quantum origins of the classical
arXiv:quant-ph/0105127v3



\end{list}



\appendix
%
\section{Stochastic Dissipative Equations of Motion}
\label{Sec:SDE}
\renewcommand{\theequation}{\Alph{section}.\arabic{equation}}
%

The reservoir or environment
introduces a stochastic dissipative element
to the time evolution of the subsystem,
$\hat{\cal U} = \hat{\mathrm I}
+ (\tau/\mathrm{i}\hbar)\hat{\overline u}^\mathrm{r}
+ \hat{\cal R}$
(Attard 2015 chapter~4).
For the expansion linear in $\tau$
this is additive to the adiabatic evolution.
The decoherent evolution of the momentum eigenfunction is taken to be
\begin{equation}
\hat{\cal U}({\bf q};\tau) \zeta_{\bf p}({\bf q})
=
 \zeta_{{\bf p}''}({\bf q}).
\end{equation}
Notice that the position is the same on both sides of this equation,
which is motivated by earlier classical analysis
(Attard 2012 sections~3.6.3 and 7.4).

The probability operator is the exponential of the entropy operator,
$\hat \wp({\bf q})
= Z^{-1} e^{\hat S^\mathrm{r}({\bf q})/k_\mathrm{B}}$,
where $k_\mathrm{B}$ is Boltzmann's constant.
In the  present canonical equilibrium case
$\hat S^\mathrm{r}({\bf q}) =- \hat {\cal H}({\bf q})/T$,
where $T$ is the temperature.
The stochastic dissipative propagator
stabilizes the probability (Attard 2015 chapter~4),
and for the decoherent subsystem this condition is
\begin{eqnarray}
\lefteqn{
\zeta_{\bf p}({\bf q})^*\,
\hat \wp({\bf q})
\zeta_{\bf p}({\bf q})
}  \\ \nonumber
& = &
\left\langle
\zeta_{\bf p}({\bf q})^*\,
\hat{\cal U}({\bf q};-\tau) \hat \wp({\bf q}) \hat{\cal U}({\bf q};-\tau)^\dag
\zeta_{\bf p}({\bf q})
\right\rangle_\mathrm{stoch} .
\end{eqnarray}

The left hand side is
\begin{eqnarray}
 \zeta_{\bf p}({\bf q})^*
 \frac{ e^{-\beta \hat {\cal H}({\bf q})} }{Z}
 \zeta_{\bf p}({\bf q})
& \equiv &
 \zeta_{\bf p}({\bf q})^* \zeta_{\bf p}({\bf q})
\frac{e^{-\beta {\cal H}({\bf q},{\bf p})}}{Z}
\omega({\bf q},{\bf p})
\nonumber \\ & = &
\frac{ e^{ S^\mathrm{eff}({\bf q},{\bf p})/k_\mathrm{B} } }{V^N Z} .
\end{eqnarray}
This defines the commutation function $\omega$
(Attard 2018, 2021, 2023a).
The effective entropy,
$S^\mathrm{eff} = -{\cal H}/T + k_\mathrm{B}\ln \omega$,
can be redefined to include wave function symmetrization,
$S^\mathrm{eff} = -{\cal H}/T + k_\mathrm{B}\ln(\omega \eta^\pm)$,
where the symmetrization function $\eta^\pm({\bf q},{\bf p})$
has been discussed in detail elsewhere
(Attard 2018, 2021, 2023a, 2023b).

The right hand side to linear order in $\tau$ is
\begin{eqnarray}
 \zeta_{\bf p}({\bf q})^* \hat{\cal U}({\bf q};-\tau)
\frac{e^{-\beta \hat{\cal H}({\bf q})}}{Z}
\lefteqn{
\hat{\cal U}({\bf q};-\tau)^\dag \zeta_{\bf p}({\bf q})
}\nonumber  \\
& = &
\zeta_{{\bf p}''}({\bf q})^*
\frac{e^{-\beta \hat{\cal H}({\bf q})}}{Z}
\zeta_{{\bf p}''}({\bf q})
\nonumber \\ & = &
\frac{ e^{ S^\mathrm{eff}({\bf q},{\bf p}'')/k_\mathrm{B} } }{V^N Z} .
\end{eqnarray}

Now  write
${\bf p}''({\bf q},{\bf p}) = {\bf p}
- |\tau| \overline {\bf u}^\mathrm{r}({\bf q},{\bf p}) + \tilde {\bf R}_p$
(Attard 2012 section~7.4.2).
The stochastic  `force' has zero mean
and diagonal variance $\sigma^2$.
Equating the two sides,
performing a second order Taylor expansion,
and stochastic averaging yields
\begin{eqnarray}
0 & = &
\frac{-|\tau|}{k_\mathrm{B}} \overline {\bf u}^\mathrm{r}
\cdot \nabla_p S^\mathrm{eff}
+ \frac{1}{2 k_\mathrm{B}^2} \langle \tilde {\bf R}_p \tilde {\bf R}_p
\rangle_\mathrm{stoch}
\nonumber \\ && \mbox{ }
: \big[ (\nabla_p S^\mathrm{eff}) (\nabla_p S^\mathrm{eff})
+ k_\mathrm{B}(\nabla_p\nabla_p S^\mathrm{eff}) \big]
\nonumber \\ & = &
\frac{-|\tau|}{k_\mathrm{B}} \overline {\bf u}^\mathrm{r}
\cdot \nabla_p S^\mathrm{eff}
+ \frac{\sigma^2}{2 k_\mathrm{B}^2}
(\nabla_p S^\mathrm{eff}) \cdot (\nabla_p S^\mathrm{eff})
\nonumber \\ && \mbox{ }
+ \frac{\sigma^2}{2 k_\mathrm{B}} \nabla_p^2 S^\mathrm{eff} .
\end{eqnarray}
The first and second terms cancel when
\begin{equation}
\overline {\bf u}^\mathrm{r}({\bf q},{\bf p})
=
\frac{\sigma^2}{2 |\tau| k_\mathrm{B}}
\nabla_p S^\mathrm{eff}({\bf q},{\bf p}).
\end{equation}
With this the final term is
\begin{equation}
\frac{\sigma^2}{2 k_\mathrm{B}} \nabla_p^2 S^\mathrm{eff}({\bf q},{\bf p})
=
|\tau| \nabla_p \cdot \overline {\bf u}^\mathrm{r}({\bf q},{\bf p}).
\end{equation}
This cancels with the compressibility of the dissipative equations of motions,
which occurs for the evolution of the probability density
in the  momentum continuum (Attard 2018, 2021, 2023a, 2023b).
For the canonical equilibrium case,
and neglecting the commutation and symmetrization functions,
the dissipative force is $\overline {\bf u}({\bf q},{\bf p})
= (-{\sigma^2}/{2 m |\tau| k_\mathrm{B}T}){\bf p}$,
which has the form of a drag or friction force
(Attard 2012 chapters~3 and 7).


\vfill

\end{document}